\pdfoutput=1
\documentclass[aps,prd,preprint,showpacs,showkeys,nofootinbib,tightenlines]{revtex4-1}

\usepackage{amsmath,amssymb}
\usepackage{bm}
\usepackage{graphicx}
\usepackage{ulem}
\usepackage{color}

\newcommand\beq{\begin{equation}}
\newcommand\eeq{\end{equation}}
\newcommand\bea{\begin{eqnarray}}
\newcommand\eea{\end{eqnarray}}

\begin{document}

\title{Point Production of a Nonrelativistic Unparticle \\Recoiling Against a Particle}

\author{Eric Braaten}
\affiliation{Department of Physics,
         The Ohio State University, Columbus, OH\ 43210, USA}

\author{Hans-Werner Hammer}
\affiliation{Technische Universit\"{a}t Darmstadt, Department of Physics, Institut f\"{u}r Kernphysik, 64289 Darmstadt, Germany}
\affiliation{ExtreMe Matter Institute EMMI and Helmholtz Forschungsakademie Hessen für FAIR (HFHF), GSI Helmholtzzentrum f\"{u}r Schwerionenforschung GmbH,
64291 Darmstadt, Germany}

\date{\today}

\begin{abstract}
A nonrelativistic unparticle can be defined as an excitation created by an operator
with a definite scaling dimension in a nonrelativistic field theory
with an approximate conformal symmetry.
The point production rate of an unparticle has power-law dependence on
its total energy with an exponent determined by its scaling dimension.
We use the exact result for the 3-point function of primary operators
in a nonrelativistic conformal field theory to derive the contribution
to the point production rate of the unparticle from its decay into
another unparticle recoiling against a particle.
In the case where the conformal symmetry is broken by a large positive
scattering length, we deduce the exponent of the energy in the
point production rate of the loosely bound two-particle state recoiling
against  a particle with large relative momentum.
\end{abstract}

\maketitle

\section{Introduction}
\label{sec:Intro}

A definition of an {\it elementary particle}
that is popular in group-theoretic circles is an irreducible representation of the Poincar\'e group.
The concept of an {\it unparticle} was introduced by Georgi \cite{Georgi:2007ek}.
An unparticle is a system created by a local operator with a definite scaling dimension in a conformal field theory (CFT).
It can therefore be defined
as an irreducible representation of the conformal symmetry group.
The conformal group on a 3+1-dimensional Minkowski space-time
is a 15-dimensional  group that includes the Poincar\'e group and scale transformations as subgroups.
A relativistic unparticle is characterized by a single number:  the scaling dimension of the operator.
If the CFT belongs to a hidden sector beyond the Standard Model of particle physics,
the unparticle cannot be observed directly.
However it can be observed indirectly through the momentum distribution of Standard Model
particles produced in association with the unparticle  \cite{Georgi:2007ek}.
If the unparticle is produced in association with a single Standard Model particle,
the invariant mass distribution of the unparticle can be determined by measuring 
the recoil-momentum distribution of the Standard Model particle.
It has power-law behavior with an exponent determined by the scaling dimension of the unparticle.
The existence of unparticles in a hidden sector would produce novel signals in high energy colliders 
\cite{Cheung:2007zza,Georgi:2007si,Cheung:2007ap}.
The CMS collaboration has searched for signals of unparticles in $pp$ collisions at the Large Hadron Collider 
\cite{Khachatryan:2014rra,Khachatryan:2015bbl,Sirunyan:2017onm}.

Hammer and Son recently pointed out that unparticles can also arise in nonrelativistic physics \cite{Hammer:2021zxb}.
The nonrelativistic conformal  symmetry group on a 3+1-dimensional Galilean  space-time  
is a  13-dimensional group that includes the Galilean group and scale transformations as subgroups.\footnote{For a discussion of the relation to the relativistic conformal group, see e.g. Ref.~\cite{Karananas:2021bqw}.}
It is also called the Schr\"odinger group, because it is the symmetry group of the free Schr\"odinger equation.
A nonrelativistic conformal field theory (NRCFT) is a field theory with  nonrelativistic conformal  symmetry \cite{Nishida:2007pj}.
A {\it nonrelativistic unparticle} is a system created by a local operator
 with a definite scaling dimension in such a theory.
 In contrast to the relativistic case, a nonrelativistic unparticle is characterized by two numbers: its kinetic mass $M$ 
 and the scaling dimension $\Delta$ of the operator \cite{Hammer:2021zxb}.

Hammer and Son pointed out that systems of low-energy neutrons produced by a short-distance reaction 
provide physical examples of nonrelativistic  unparticles \cite{Hammer:2021zxb}.
Neutrons have a negative scattering length $a$ that is much larger than their effective range $r_0$.
As a consequence, the behavior of a system of neutrons whose kinetic energies in their center-of-momentum (CM) frame
are all in the scaling region between $1/(m_n a^2)$ and $1/(m_n r_0^2)$,
where $m_n$ is the neutron mass, is approximately scale invariant.
In the unitary limit $1/a \to 0$, $r_0 \to 0$,
a nonrelativistic effective field theory describing the neutrons has nonrelativistic conformal symmetry.
A convenient basis for local operators in the effective field theory are those with definite scaling behavior under the conformal symmetry.
A system of $N$ neutrons created by a local operator with scaling dimension $\Delta_N$
can be interpreted as an unparticle with mass $Nm_n$.
For the 2-neutron unparticle, the lowest scaling dimension is $\Delta_2=2$.
For unparticles with larger $N$, the scaling dimensions are transcendental numbers.
For the 3-neutron unparticle, the lowest scaling dimension is $\Delta_3=4.27272$.
The lowest scaling dimensions for 4 and 5 neutrons are 5.07 and 7.6, respectively.
 A summary of the lowest scaling dimensions for up to 6 neutrons is
given in Table~I of \cite{Nishida:2007pj}.

\begin{figure}[t]
\includegraphics[width=0.75\textwidth]{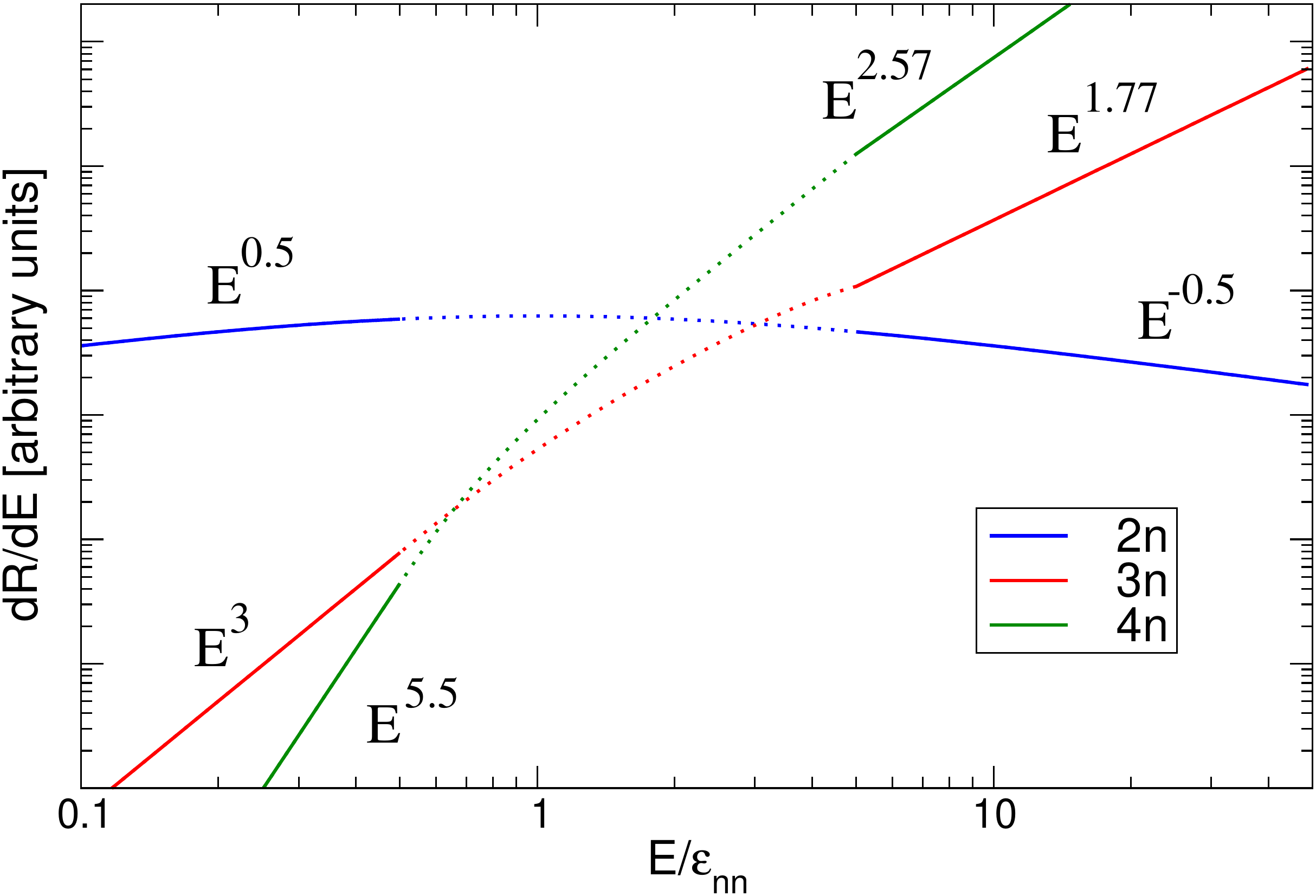}
\caption{
Scaling prediction of  Ref.~\cite{Hammer:2021zxb} for the invariant energy
distribution $dR/dE$
of 2, 3, and 4 neutrons in their CM frame. The transition from
the free phase-space behavior to the scaling region around $E\sim
\varepsilon_{nn}$ is illustrated by the dotted lines.
}
\label{fig:neutrons}
\end{figure}

An $N$-neutron unparticle can be created 
by a short-distance nuclear reaction of the form $A_1+A_2 \longrightarrow B+(nn\ldots)$   \cite{Hammer:2021zxb}.
The invariant energy $E$  of the $N$ neutrons, 
which is their total kinetic energy in their CM frame,
can be determined by measuring the momentum of the recoiling nucleus $B$
or by detecting the neutrons directly.
There is a scaling region of $E$ 
in which the differential cross section has the scaling behavior $E^{\Delta_N-5/2}dE$.
The corresponding power-law behavior for $N$ noninteracting particles as $E$ approaches the threshold 
is governed by the $N$-particle phase space, $E^{(3N-5)/2}dE$.
In the case of fermions, there is an additional Pauli suppression factor
$E$ for each pair of identical particles.
For $E$ of the order of the energy scale set by the
neutron-neutron scattering length $\varepsilon_{nn}=1/(m_n a^2)$,
there is a transition from the phase-space
behavior for non-interacting particles to the power-law 
behavior governed by $\Delta_N$ in the scaling region. This nontrivial
scaling behavior is the smoking gun for an unparticle.
The resulting predictions for the invariant energy distributions
of 2, 3, and 4 neutrons are shown in Fig.~\ref{fig:neutrons}.
In the case of 2 neutrons, there is a maximum in the region
$E\sim\varepsilon_{nn}$. For larger neutron
numbers, there is a change in the slope
of the log-log plot in the region $E\sim\varepsilon_{nn}$.

The predicted invariant energy distribution, $dR/dE$, for four neutrons
in Fig.~\ref{fig:neutrons} is not consistent with a resonance-like
structure at
$E/\varepsilon_{nn}\approx 20$ in the point production of four neutrons.
Such a structure was recently observed in the $4n$ spectrum measured in the
knock-out reaction $^8$He$(p,p\alpha)4n$ by Duer et al.~\cite{Duer:2022ehf}.
However, the neutron distribution of the initial $^8$He nucleus clearly
plays a role in this process and the applicability of the point-production
approximation is questionable as the neutrons are emitted from a $^8$He source.
A recent theoretical study of the
reaction~\cite{Lazauskas:2022mvq} attributes the structure to the final state interaction among the
four neutrons and the presence of preexisting four neutrons in the periphery
of $^8$He nucleus projectile. While the former effect is captured
in the  the invariant $4n$ energy distribution of Fig.~\ref{fig:neutrons},
the latter is not.

In Ref.~\cite{Braaten:2021iot},
we pointed out that a low-energy system of neutral charm mesons created by a short-distance reaction 
is an unparticle because of the $X(3872)$ resonance in the $D^{*0} \bar D^0 + D^0 \bar D^{*0}$ channel \cite{Choi:2003ue}.
The scattering length $a$ in that channel is large and negative.
Because the resonant channel is a superposition of charm mesons, there is no Efimov effect \cite{Braaten:2003he,Braaten:2004rn}.
The behavior of a system of neutral charm mesons whose kinetic energies in their CM frame
are all in the scaling region between $1/(2\mu a^2)$ and  $1/(2\mu r_0^2)$,
where $\mu$ is the reduced mass of $D^{*0}$ and $\bar D^0$, is therefore approximately scale invariant.
In the unitary limit $1/a \to 0$, $r_0 \to 0$,  
a nonrelativistic effective field theory describing the neutral charm mesons has nonrelativistic conformal symmetry.
A convenient basis for local operators in the effective field theory are those with definite scaling behavior under the conformal symmetry.
A system of $N$ neutral charm mesons created by a local operator with scaling dimension $\Delta_N$ 
can be interpreted as an unparticle.
The scaling dimensions for 2-charm-meson unparticles are the same as for two neutrons.
The lowest  scaling dimensions for 3-charm-meson unparticles
are $\Delta_3= 3.10119$ for $D^0 \bar{D}^{*0} D^0$ and
$\bar{D}^0 D^{*0} \bar{D}^0$ as well as
$\Delta_3= 3.08697$ for $D^0 \bar{D}^{*0} \bar{D}^{*0}$ and
$\bar{D}^0 D^{*0} D^{*0}$.
In each case, there are two pairs of particles with a large scattering
length $a$. The slight difference in the two values of $\Delta_3$ is due to
the different masses of the $D^0$ and $D^{*0}$.
The 2-charm-meson unparticle can be observed through the recoil-momentum  spectrum 
of the kaon in the inclusive decay $B^\pm \to K^\pm +(\mathrm{anything})$.
A 3-charm-meson unparticle can be observed through the prompt production of $X(3872)\, D^0$
at the Large Hadron Collider.

Because the scattering length $a$ in the case of neutral charm mesons  is
large and positive, 
the unparticles have components that include bound states.
There are reactions for producing final states that include bound states whose rates have power-law behavior 
with exponents determined by conformal symmetry.
The simplest such reaction is the production of $X(3872)\, D^0$ from the creation of 
a 3-charm-meson unparticle  at a point, whose rate scales as a power of the unparticle energy $E$~\cite{Braaten:2021iot}.

In this paper, we consider  the production of a pair of unparticles from  the creation of a single unparticle at a point.
We use the analytic result for the 3-point function of primary operators
in a NRCFT to derive the contribution
to the production rate from another unparticle recoiling against a single particle.
In the case where the conformal symmetry is broken by a large positive scattering length,
we deduce the exponent of the energy in the point production rate 
of the loosely bound two-particle state recoiling against  a particle
with large relative momentum.

\section{Three-point function for primary operators}
\label{sec:LoopInt}

We start by 
calculating the Fourier transform of the 3-point function for primary operators in a NRCFT.
We take the dimension of space to be $D$, and we denote a space-time position by $x=(\bm{x},t)$.
An operator $\phi_3^\dagger(x)$ with scaling dimension $\Delta_3$ creates an unparticle with mass $M_3$. 
An operator $\phi_2(x)$ with scaling dimension $\Delta_2$ annihilates an unparticle with mass $M_2$.
An operator $\phi_1(x)$ with scaling dimension $\Delta_1$ annihilates an unparticle with mass $M_1=M$.
The masses satisfy the constraint $M_1+M_2=M_3$ from Galilean symmetry.
We are particularly interested in the case
where $\phi_1(x)$ annihilates a single particle with mass $M$, 
in which case its scaling dimension is $\Delta_1=D/2$.

\subsection{Propagators}

\begin{figure}[t]
\includegraphics[width=0.3\textwidth]{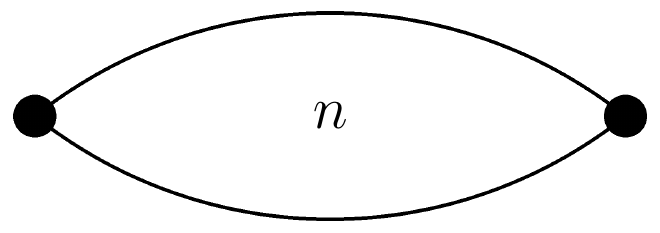}
\caption{
Space-time propagator for unparticle $n$ with kinetic mass $M_n$ and scaling dimension $\Delta_n$.
}
\label{fig:propagator}
\end{figure}

The space-time propagator for a primary operator $\phi_n(x)$  with mass $M_n$ and scaling dimension $\Delta_n$  is
\beq
\big\langle  \phi_n (x_1) \, \phi_n^\dagger (x_2) \big\rangle =
C_n\, (t_{12})^{-\Delta_n}  \theta(t_{12}) 
 \, \exp\! \bigg( i  \frac{M_n \, x_{12}^2}{2\,  t_{12}}  \bigg),
\label{propn-xt}
\eeq
where $t_{ij} = t_i-t_j$, $x_{ij}^2 = (\bm{x}_i - \bm{x}_j)^2$, and $C_n$ is a constant.
We represent the space-time propagator by the diagram in
Fig.~\ref{fig:propagator}.

The energy-momentum propagator is obtained by Fourier-transforming in both space-time positions
and then factoring out an  energy-momentum  delta function.
It is advantageous to take the variables $(x_1+x_2)/2$ and $x_1-x_2$
for the Fourier transform. After performing the trivial Fourier transform
in $(x_1+x_2)/2$, the
remaining Fourier transform in space is a Gaussian integral that depends on the momentum $\bm{p}$:
\beq
\int d^Dx_{12}  \, \exp\! \bigg(\! -i \,  \bm{p}\cdot \bm{x}_{12} +i \frac{M_n\, x_{12}^2}{2  t_{12}}   \bigg) =
\left( 2 \pi i \frac{ t_{12}}{M_n} \right)^{D/2}  \, \exp\! \bigg(\!  -i  \, \frac{t_{12}}{2 M_n}  p^2  \bigg) .
\label{Fourier-x}
\eeq
The subsequent Fourier transform in time gives a simple function of the energy $E$:
\beq
\int_0^\infty \!\!\!dt_{12} \, ( i t_{12})^{D/2 -\Delta_n} 
 \, \exp\! \bigg(\! i \, E \, t_{12} -i  \, \frac{  p^2}{2 M_n}  t_{12} \! \bigg) =- i\,
\Gamma \bigg(\! \frac{D}{2} + 1 - \Delta_n \! \bigg)  \left(  \frac{p^2}{2M_n} - E \right)^{\Delta_n - D/2  - 1}.
\label{Fourier-t}
\eeq
With an appropriate choice for the constant $C_n$ in Eq.~\eqref{propn-xt},
our final result for the energy-momentum propagator is 
\beq
D_n(E,p) = -i \, C_n^\prime
\left(  \frac{p^2}{2M_n} - E \right)^{\Delta_n-D/2 - 1},
\label{propn-Ep}
\eeq
where  $C_n^\prime$ is a constant.

In the case where $\phi_1(x)$ annihilates a single particle,  its scaling dimension is $\Delta_1=D/2$
and the propagator in Eq.~\eqref{propn-Ep} has a simple pole in the energy $E$ at $p^2/(2M)$.
With the conventional choice $C_1^\prime = 1$, the discontinuity in the energy is
\beq
D_1(E+ i \epsilon,p) - D_1(E- i \epsilon,p) = 2\pi\,  \delta\bigg(\! E -  \frac{p^2}{2M} \bigg).
\label{Discprop1}
\eeq
For other unparticles, the discontinuity in the energy is
\beq
D_n(E+ i \epsilon,p)\! -\! D_n(E- i \epsilon,p) = 2C_n^\prime\, 
\sin\big( \pi (\Delta_n\! -\! D/2)\big)
\left(\!  E\! - \! \frac{p^2}{2M_n} \right)^{\Delta_n  -D/2 - 1} \theta\bigg(\! E\! -\!  \frac{p^2}{2M_n} \bigg) .
\label{Discpropn}
\eeq

\subsection{Space-time three-point function}

\begin{figure}[t]
\includegraphics[width=0.4\textwidth]{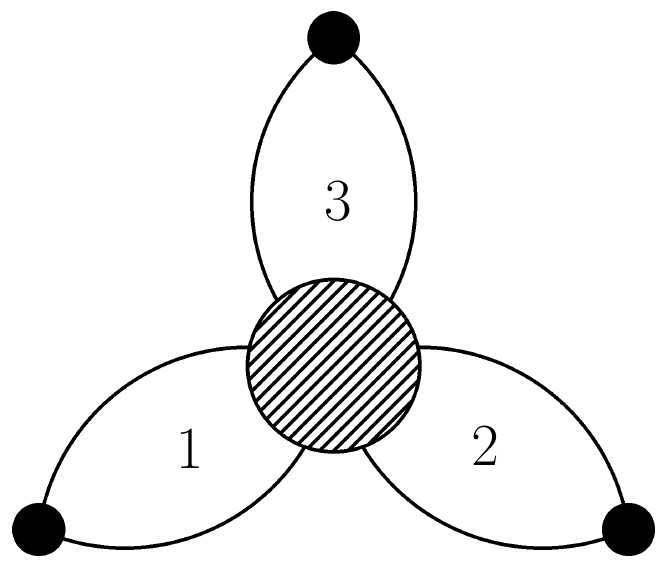}
\caption{
Space-time 3-point function for primary operators associated with unparticles 1, 2, and 3.
}
\label{fig:3ptfunction}
\end{figure}

The 3-point function $\big\langle  \phi_1(x_1)\, \phi_2(x_2) \, \phi_3^\dagger  (x_3)  \big\rangle$
for primary operators in a NRCFT was first considered by Henkel in 1993 \cite{Henkel:1993sg}.
The 3-point function can be represented by the diagram in Fig.~\ref{fig:3ptfunction}.
He  used the Schr\"odinger symmetry  to determine the 3-point function analytically
up to a scaling function of a single variable:
\bea
\big\langle  \phi_1(x_1)\, \phi_2(x_2) \, \phi_3^\dagger  (x_3)  \big\rangle &=& 
(t_{13})^{-\Delta_{13,2}/2}\theta(t_{13})\, \exp\! \bigg(\! i \frac{M_1 \, x_{13}^2}{2\,  t_{13} } \bigg)
 \nonumber\\
 &&\times
(t_{23})^{-\Delta_{23,1}/2}\theta(t_{23})\, \exp\! \bigg(\! i  \frac{M_2 \, x_{23}^2}{2\, t_{23}} \bigg)\,
(t_{12})^{-\Delta_{12,3}/2} \, \Phi (w),
\label{<phi123>tx}
\eea
where $\Delta_{ij,k} = \Delta_i + \Delta_j - \Delta_k$ and the argument of the scaling function $\Phi$ is 
\beq
w  =  \frac{\big( t_{23} (\bm{x}_1 - \bm{x}_3) - t_{13} (\bm{x}_2 - \bm{x}_3) \big)^2}{2\, t_{13} \, t_{23}\,  t_{12}}
=  \frac12 \left( \frac{x_{12}^2}{t_{12}} - \frac{x_{13}^2}{t_{13}} +  \frac{x_{23}^2}{t_{23}} \right).
\label{w123}
\eeq
The first expression shows that the sign of $w$ is the same as the sign of $t_{12}$.
The scaling function $\Phi(w)$ depends also on the masses $M_1$ and $M_2$
and on the scaling dimensions $\Delta_1$, $\Delta_2$, and $\Delta_3$.
The 3-point function in Eq.~\eqref{<phi123>tx} is symmetric under the interchange of the subscripts 1 and 2.

An integral representation for the scaling function $\Phi(w)$ was first obtained by Henkel and Unterberger
from the 3-point function for a  CFT in two higher dimensions \cite{Henkel:2003pu}.
It has also been obtained by Fuertes and Moroz and by Volovich and Wen
using holography and the AdS/CFT correspondence \cite{Fuertes:2009ex,Volovich:2009yh}.
In the Anti-de Sitter space (AdS) formulation, the external points in Fig.~\ref{fig:3ptfunction}
are on the boundary of AdS and the blob is replaced by a point interaction inside the bulk of AdS.
In Refs.~\cite{Henkel:2003pu,Fuertes:2009ex}, the 3-point function is expressed as a function of Euclidean times.
The expression in Ref.~\cite{Volovich:2009yh} in terms of real times is more directly useful for our purposes.
The integral representation for the scaling function is
\bea
\Phi(w) &=& 
C_{12,3} \,\int_{-\infty}^{+\infty} du \, (u + i \epsilon)^{-\Delta_{13,2}/2} \, e^{-i M_1 u} 
\int_{-\infty}^{+\infty} dv \, (v+ i \epsilon)^{-\Delta_{23,1}/2}\, e^{- iM_2 v}\,
 \nonumber\\
 &&
 \hspace{2cm} \times \big[ u-v+ (1+i\epsilon) w \big]^{-\Delta_{12,3}/2},
\label{Phi-w}
\eea
where $C_{12,3}$ is  a constant.
Because the signs of $t_{12}$ and $w$ are identical (cf.~Eq.\eqref{w123}),
the product of $(t_{12})^{-\Delta_{12,3}/2}$ and the last factor in Eq.~\eqref{Phi-w} has a well-defined imaginary part:
\beq
(t_{12})^{-\Delta_{12,3}/2} \, \big[ u-v+ (1+i\epsilon) w \big]^{-\Delta_{12,3}/2} =
\big[ t_{12}\,  (u-v+ w) + i\epsilon \big]^{-\Delta_{12,3}/2}.
\label{t(u-v+w)}
\eeq
The scaling function can also be  expressed analytically in terms of a confluent hypergeometric function \cite{Volovich:2009yh}.

\subsection{Fourier transform in space}

\begin{figure}[t]
\includegraphics[width=0.4\textwidth]{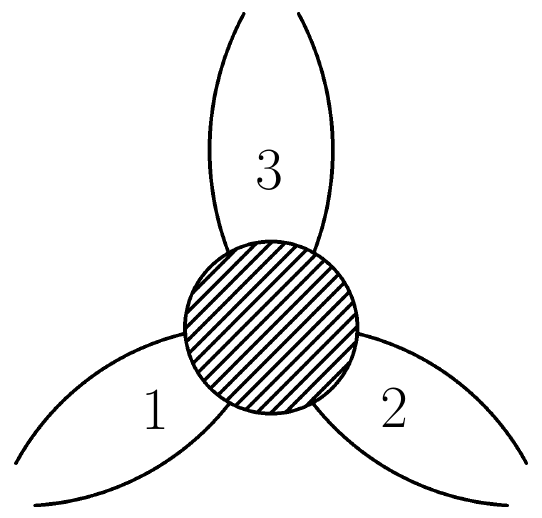}
\caption{
Fourier-transformed 3-point function for primary operators associated with unparticles 1, 2, and 3.
}
\label{fig:3ptfunctionFT}
\end{figure}

The Fourier transform of the 3-point function in position space
can be represented by the diagram in Fig.~\ref{fig:3ptfunction}.
To evaluate the Fourier transforms in  the three spatial positions,
it is convenient to express the last factor in Eq.~\eqref{t(u-v+w)} as the integral of an exponential:
\bea
\big[ t_{12} (u-v+ w)+i\epsilon \big]^{-\Delta_{12,3}/2} &=& 
\big( e^{i \pi/2} \, |t_{12}| \big)^{-\Delta_{12,3}/2} \frac{1}{\Gamma(\Delta_{12,3}/2)}
\nonumber\\ 
&& \hspace{-2cm}
\times \int _0^\infty\!\!\! dm \,  m^{\Delta_{12,3}/2-1} \exp \big( -m [\epsilon -i \, \mathrm{sign}(t_{12}) (u-v+ w)] \big).
\label{uvwint}
\eea
The Fourier transform can be expressed as the product 
of the  momentum-conserving delta function $\delta^D(\bm{p}_1 + \bm{p}_2 - \bm{p}_3)$
and a Gaussian integral in $\bm{x}_{13}$ and $\bm{x}_{23}$.
In the case $t_{12}>0$, the Gaussian integral is
\bea
&&\int d^Dx_{13}  \, \exp\! \bigg(\! -i \,  \bm{p}_1\cdot \bm{x}_{13} +i \frac{M_1\, x_{13}^2}{2  t_{13}}   \bigg) 
\int d^Dx_{23}  \, \exp\! \bigg(\! -i \,  \bm{p}_2\cdot \bm{x}_{23} +i \frac{M_2\, x_{23}^2}{2  t_{23}}   \bigg) \,
e^{+i \, m \, w}
\nonumber\\ 
&&=
\left( 2 \pi i \,\frac{ t_{12}  t_{13}}{ t_{12} M_1+ t_{23}m} \right)^{D/2}  \, 
\left( 2 \pi i \,\frac{ t_{23} ( t_{12} M_1+ t_{23}m)}{t_{12}M_1M_2 + t_{13}M_1 m + t_{23}M_2 m} \right)^{D/2}  \, 
\nonumber\\ 
&& \hspace{1cm}
\times \exp\! \bigg(\! -i  \frac{ t_{12} t_{13}(M_2+m)p_1^2 +  t_{12} t_{23}(M_1-m)p_2^2 + t_{13}t_{23} m p_3^2 }
{2 (t_{12}M_1M_2 + t_{13}M_1 m + t_{23}M_2 m) }  \bigg).
\label{intx12x13}
\eea

We can go to the center-of-momentum (CM) frame by setting $\bm{p}_3=0$ and $|\bm{p}_1| = |\bm{p}_2|= p$.
In the case $t_{12}>0$, the 3-point function with the momentum delta function factored out reduces to
\bea
&& C_{12,3}^\prime\, 
(t_{13})^{(D-\Delta_{13,2})/2}\theta(t_{13})  \,(t_{23})^{(D-\Delta_{23,1})/2} \theta(t_{23})  \,
|t_{12}|^{-\Delta_{12,3}/2}
\int _0^\infty\!\!\!dm  \, m^{\Delta_{12,3}/2-1}\,  e^{-m \epsilon}
 \nonumber\\
 &&\times
 \int_{-\infty}^{+\infty} du \,  (u + i \epsilon)^{-\Delta_{13,2}/2} \, e^{-i (M_1-m) u} 
\int_{-\infty}^{+\infty} dv \,  (v + i \epsilon)^{-\Delta_{23,1}/2}\, e^{- i (M_2+m) v}\,
 \nonumber\\
 &&\times
 \left( \frac{ t_{12}M_1M_2 }{t_{12}M_1M_2 + t_{13}M_1 m + t_{23}M_2 m } \right)^{D/2}
\exp\! \bigg(\! - i \frac{t_{12}(t_{23}M_1 + t_{13} M_2 + t_{12} m) }{2 (t_{12}M_1M_2 + t_{13}M_1 m + t_{23}M_2 m) }  p^2\! \bigg),
 \nonumber\\
\label{Gaussian}
\eea
where $C_{12,3}^\prime$ is  a constant.
The integrals over $u$ and $v$ can be evaluated analytically by closing the integration contour in the lower half-plane:
\beq
 \int_{-\infty}^{+\infty} du \,  (u + i \epsilon)^{-\Delta/2} \, e^{-i M u} 
 =e^{-i\pi\Delta/4} \frac{2\pi}{\Gamma(\Delta/2)} M^{\Delta/2-1} \theta(M).
\label{intu}
\eeq
After these integrals have been evaluated, the limit $\epsilon\to 0$ can be taken in the remainder.
The choice $t_{12}>0$ breaks the symmetry under interchange of the subscripts 1 and 2.
Given this choice, it is convenient to change to a dimensionless integration variable $x= m/M_1$.
Our final result for the spatial Fourier transform of the 3-point function in the CM frame with $t_{12}>0$ is
\bea
&& C_{12,3}^{\prime \prime}\, 
(t_{13})^{(D-\Delta_{13,2})/2}\theta(t_{13})  \,  (t_{23})^{(D-\Delta_{23,1})/2}\theta(t_{23}) \, |t_{12}|^{-\Delta_{12,3}/2}
 \nonumber\\
 &&\times
\int _0^{1}dx  \, x^{\Delta_{12,3}/2-1}\,  (1-x)^{\Delta_{13,2}/2-1}\,  \big[  1 + (M_1/M_2)  x \big]^{\Delta_{23,1}/2-1}\, 
 \nonumber\\
 && \hspace{0.5cm}\times
 \left( \frac{ t_{12} }{t_{12} + (M_1/M_2) x t_{13} +  x t_{23}} \right)^{D/2}
\exp\! \bigg(\! - i \frac{t_{12} [ x t_{12}  + (M_2/M_1)t_{13} + t_{23} ]}{2 M_2 [t_{12} + (M_1/M_2) x t_{13} +  x t_{23}]}  p^2\! \bigg),
\label{<phi123>tp}
\eea
where $C_{12,3}^{\prime \prime}$ is  a constant.

\subsection{Fourier transform in time}

The Fourier transforms in the three  times  can be expressed as the product 
of the  energy-conserving delta function $\delta(E_1 + E_2-E_3)$,
a Fourier transform in $t_{13}$ with energy $E_1$, and a Fourier transform in $t_{23}$  with energy $E_2$.
We first isolate the contribution from the Fourier transform in  $t_{13}$  from the region  $t_{13}  \gg  t_{23}$ 
and then evaluate the Fourier transform in  $t_{23}$.

In  the limit  $t_{13}  \gg  t_{23}$, the exponential in Eq.~\eqref{<phi123>tp}, reduces to
\bea
&&
\exp\! \bigg(\! - i \frac{t_{12} \big[  x t_{12}  + (M_2/M_1)t_{13} + t_{23} \big] }{2 M_2 \big[t_{12} + (M_1/M_2) x t_{13} +  x t_{23} \big] }  p^2\! \bigg)
 \nonumber\\
 &&\hspace{2cm}
\longrightarrow \exp\! \bigg( - i \frac{p^2}{2M_1}  t_{13}\bigg)
\exp\! \bigg( -i  \frac{ [1- 2x - (M_2/M_1)x] \,p^2}{2M_2[ 1 + (M_1/M_2)x ] }  t_{23}\bigg) ,
\label{exp-t13,t23}
\eea
where the identity $t_{12}=t_{13}-t_{23}$ has been used.
The contribution to the Fourier transform in  $t_{13}$  from the region  $t_{13}  \gg  t_{23}$
can be reduced to the integral
\bea
&& \int_0^\infty dt_{13} \,  (i t_{13})^{(D-\Delta_{13,2}-\Delta_{12,3})/2}
 \exp\! \bigg(  i E_1  t_{13} - i \frac{p^2}{2M_1} t_{13} \bigg)
 \nonumber\\
&&\hspace{1cm}
 = -i\, \Gamma\big( D/2+1 -\Delta_1 \big)  \left( \frac{p^2}{2M_1} - E_1 \right)^{\Delta_1- D/2 -1} .
\label{FTt13} 
\eea
This integral is the propagator for $\phi_1$ multiplied by a constant.
The subsequent Fourier transform in $t_{23}$ can be reduced to the  integral
\bea
&& \int_0^\infty dt_{23} \,  (i t_{23})^{(D-\Delta_{23,1})/2}
\exp\! \bigg( i  E_2  t_{23}  -i  \frac{ [1-2x - (M_2/M_1)x] \,p^2}{2M_2[ 1 + (M_1/M_2)x ] }  t_{23} \bigg)
 \nonumber\\
&&\hspace{1cm}
 = -i\, \Gamma\big( [D-\Delta_{23,1}]/2 +1\big)  
 \left(  \frac{ [1- 2x - (M_2/M_1)x] \,p^2}{2M_2[ 1 + (M_1/M_2)x ] }  - E_2 \right)^{(\Delta_{23,1} -D)/2-1} .
\label{FTt23} 
\eea

We now specialize to the case of $\phi_1(x)$ being the operator that
annihilates a single particle. Its scaling dimension is $\Delta_1 = D/2$,
so the integral in Eq.~\eqref{FTt13} has a simple pole at $E_1 = p^2/(2M_1)$.
The residue of the pole in the Fourier-transformed three-point function is 
\bea
G_\mathrm{pole}(E_2,p) &=& C_{12,3}^{\prime \prime \prime}
\int _0^{1}dx  \, x^{\Delta_{12,3}/2-1}\,  (1-x)^{\Delta_{13,2}/2-1}\,  \big[ 1 + (M_1/M_2) x \big]^{\Delta_{23,1}/2-1-D/2}\, 
\nonumber\\
 &&  \hspace{2cm} \times
  \left( \frac{p^2}{2M_2} - E_2  -  \frac{(M_3/M_2) x}{1+ (M_1/M_2) x} \, \frac{p^2}{2M_{12}} \right)^{(\Delta_{23,1} -D)/2 - 1} ,
\label{<phi123>Ep}
\eea
where $M_{12}= M_1M_2/M_3$ is a reduced mass and $C_{12,3}^{\prime \prime  \prime}$ is  a constant.
The integrand has been expressed as a function of 
the energy $E_2-p^2/(2M_2)$ of the unparticle relative  to its threshold 
and the total energy $p^2/(2M_{12})$ of the unparticle at its threshold and the particle.

In the limit $E_2\to p^2/(2M_2)$ with $p^2$ fixed,
the integral over $x$ in Eq.~\eqref{<phi123>Ep} has a divergent term that comes from the region near the lower endpoint.
The divergent factor is the propagator of the unparticle. 
The limiting behavior of the 3-point function in Eq.~\eqref{<phi123>Ep} as $E_2\to p^2/(2M_2)$  is
\bea
G_\mathrm{pole}(E_2,p) &\longrightarrow & C_{12,3}^{\prime \prime \prime \prime} \,  D_2(E_2,p)
 \left(  \frac{p^2}{2M_{12}}  \right)^{- \Delta_{12,3}/2}, 
\label{<phi123>threshold}
\eea
where $C_{12,3}^{\prime \prime \prime \prime}$ is  a constant.  
This is our result for the scaling behavior of the Fourier-transformed 3-point function at large momentum $p$.

The exponent $- \Delta_{12,3}/2$ in Eq.~\eqref{<phi123>threshold}
can also be deduced from the analytic result for an integral 
whose integrand has the same dependence on $x$ near the lower endpoint
as the integrand in Eq.~\eqref{<phi123>Ep}:
\bea
&&\int _0^1\!\!\!dx  \, x^{\Delta_{12,3}/2-1}\,  (1-x)^{\Delta_{13,2}/2-1}\, 
  \left( \frac{p^2}{2M_2} - E_2  - x  (M_3/M_2)\frac{p^2}{2M_{12}} \right)^{(\Delta_{23,1} -D)/2 - 1} 
 \nonumber\\
&&  \hspace{1cm}
=  B\bigg(\!\frac{\Delta_{12,3}}{2},\frac{\Delta_{13,2}}{2} \!\bigg) \, 
 \left( \frac{p^2}{2M_2} - E_2 \right)^{(\Delta_{23,1} -D)/2 - 1}
 \nonumber\\
&&  \hspace{3cm} \times
  \,{}_2F_1 \bigg(\frac{\Delta_{12,3}}{2}, \frac{D+2- \Delta_{23,1}}{2} ;
  \Delta_1 ; \frac{(M_3/M_2) \,  (p^2/2M_{12})}{p^2/(2M_2) - E_2 } \bigg) ,
\label{<phi123>disc}
\eea
where $B(x,y)$ is the beta function and $_2 F_1(a,b;c;z)$ is the ordinary
hypergeometric function (cf.~Ref.\cite{Gradshteyn:1965}).
Using the transformation formula  that changes the argument of the hypergeometric function from $z$ to $1/z$,
we find that its leading divergence as $ z \to \infty$ has the factor  $z^{- \Delta_{12,3}/2}$ provided $D>2$.

\section{Point production rates}
\label{sec:Prodpartbound}

We are now in the position to determine the energy dependence of
the production rate of an unparticle recoiling against a particle with
large relative momentum
from the creation of an unparticle at short distances.
We then determine the exponent of the energy in the production rate of a loosely bound 2-particle state recoiling against a particle with large relative momentum.
For definiteness, we focus on the case $D=3$.

\subsection{Unparticle recoiling against a particle}
\label{sec:pup}
\begin{figure}[t]
\includegraphics[width=0.4\textwidth]{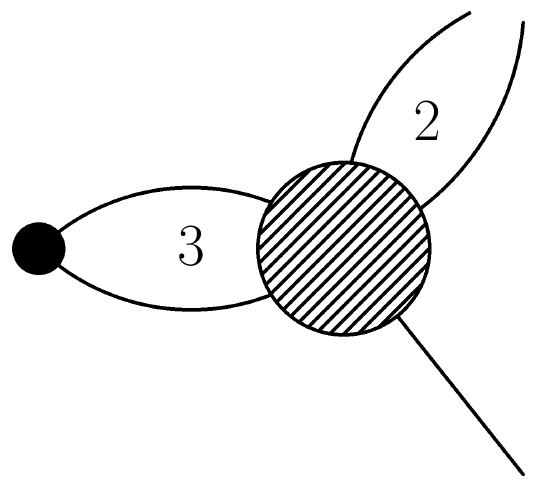}
\caption{
Amplitude for the creation of unparticle 3 at a point 
and its evolution to unparticle 2 and a single particle with large relative momentum.
The particle is represented by a single line.
}
\label{fig:pt3to2+1}
\end{figure}

We first consider the production rate in the NRCFT of 
a particle with mass $M_1$  plus the unparticle with mass $M_2$ and scaling dimension $\Delta_2$ 
from the creation at short distances of the unparticle with mass $M_3 = M_2+M_1$ and scaling dimension $\Delta_3$.
The inclusive production rate is proportional to the imaginary part of the Green function 
$\big\langle  \phi_3 (x_1) \, \phi_3^\dagger (x_2) \big\rangle$.
The contribution to the production rate 
from  intermediate states consisting of a particle and an unparticle
can be obtained from the Fourier transform of the 3-point function 
$\big\langle  \phi_1(x_1)\, \phi_2(x_2) \, \phi_3^\dagger  (x_3)  \big\rangle$
in the CM frame, which we denote as $G(E_1,E_2,p)$.
The Green function $G(E_1,E_2,p)$ with the particle propagator
$D_1(E_1,p)$ amputated
and evaluated on shell at $E_1 =p^2/(2M)$ is $G_\mathrm{pole}(E_2,p)$
in Eq.~\eqref{<phi123>Ep}.
That amplitude can be represented by the diagram in Fig.~\ref{fig:pt3to2+1}.
By energy conservation, the total energy is $E_3 = E_1+E_2$.
Using the optical theorem,
the differential rate $dR$ as a function of the total energy $E_3$ 
can be obtained from $G_\mathrm{pole}(E_2,p)$ in several steps: 
\begin{itemize}
\item
complete the amputation of the final-state propagators by dividing
$G_\mathrm{pole}(E_2,p)$ by the unparticle propagator $D_2(E_2,p)$,
\item
multiply the amputated amplitude
$G_\mathrm{amp}(E_2,p) = G_\mathrm{pole}(E_2,p)/D_2(E_2,p)$
 by its complex conjugate, 
\item
multiply  $|G_\mathrm{amp}(E_2,p)|^2$ by the discontinuities in the particle propagator $D_1(E_1,p)$ 
and in the unparticle propagator $D_2(E_2,p)$, 
which are given in Eqs.~\eqref{Discprop1} and \eqref{Discpropn},
\item
 integrate over the energy $E_1$ of the particle and its momentum $\bm{p}$ in the CM frame
 with the measure  $dE_1 d^3p/(2\pi)^4$.
\end{itemize}
The resulting expression for the differential production rate in the CM frame as a function of $E_3$ is 
\bea
dR &=& C \int \frac{d E_1}{2\pi}  \frac{d^3p}{(2\pi)^3} \,
\big| G_\mathrm{amp}(E_3-E_1,p)\big|^2\,
2\pi\,  \delta\bigg(\! E_1 -  \frac{p^2}{2M_1}\bigg) 
 \nonumber\\
 && \hspace{3cm}  \times 
 \left(\!  E_3-E_1- \! \frac{p^2}{2M_2} \right)^{\Delta_2  -5/2} \theta\bigg(\! E_3-E_1 -\!  \frac{p^2}{2M_2} \bigg),
\label{dR-threshold1}
\eea
where  $C$ is  a constant.
The integral over $E_1$ can be evaluated using the delta function, which sets $E_1=p^2/(2M_1)$.
The discontinuity in  $D_2(E_2,p)$ provides the threshold $E_3 > p^2/(2M_{12})$, 
where $M_{12}= M_1M_2/M_3$ is a reduced mass.
The differential rate for $E_3$ above the threshold reduces to
\beq
dR = C \left(\!  E_3 - \! \frac{p^2}{2M_{12}} \right)^{\Delta_2  -5/2}
\big| G_\mathrm{amp}(E_3-p^2/(2M_1),p)\big|^2 \,  \frac{d^3p}{(2\pi)^3}.
\label{dR-threshold2}
\eeq

We can use Eq.~\eqref{<phi123>threshold} to deduce 
the rate for energy $E_3$ close to the threshold  $p^2/(2M_{12})$ for fixed $p$.
The limiting behavior  as $E_2 \to p^2/(2M_2)$ of the amputated 3-point
function is
\bea
G_\mathrm{amp}(E_2,p) &\longrightarrow & C_{12,3}^{\prime \prime \prime \prime} \, 
 \left(  \frac{p^2}{2M_{12}}  \right)^{- \Delta_{12,3}/2}.
\label{Gampthreshold}
\eea
The limiting behavior of the differential rate as $E_3 \to p^2/(2M_{12})$ is therefore
\beq
dR \longrightarrow C^{\prime} \left(  E_3 - \frac{p^2}{2M_{12}} \right)^{\Delta_2-5/2}
\left( \frac{p^2}{2M_{12}} \right)^{\Delta_{3}-\Delta_2-3/2}  \, \frac{d^3p}{(2\pi)^3},
\label{dR-threshold3b}
\eeq
where $C^\prime = C_{12,3}^{\prime \prime \prime \prime} C$ and we have used
$\Delta_1=3/2$.

The above results for the point production rate in Eq.~\eqref{dR-threshold3b} can be applied to multi-neutron systems.
In this case,  $\Delta_3$ and  $\Delta_2$ are the scaling dimensions of $(N+1)$-neutron 
and $N$-neutron operators, respectively.
The contribution to the point production rate of the $(N+1)$-neutron unparticle from its decay into 
an $N$-neutron unparticle recoiling against a single neutron is given by Eq.~\eqref{dR-threshold2}.
In the region near the threshold for the $N$-neutron unparticle, that production rate reduces
at large relative momentum to Eq.~\eqref{dR-threshold3b}.
The lowest scaling dimensions for 3, 4, and 5 neutrons 
are 4.27, 5.07, 7.6, respectively. 
(See \cite{Nishida:2007pj} for further discussion and references.)

\subsection{Bound state  recoiling against a particle}
\label{sec:pbs}
Next we consider the case  where
a bound state arises from a deformation of a nonrelativistic conformal
field theory
that produces a large positive scattering length $a$
in the channel associated with the unparticle with mass $M_2$ and scaling dimension $\Delta_2 = 2$.
The bound state has constituents with masses $M_1$ and $M_2-M_1$.
Its binding energy is $|\varepsilon_X| = 1/(2 \mu a^2)$,
where $\mu = M_1(M_2- M_1)/M_2$ is the reduced mass of its constituents. 
The field $\phi_2^\dagger(x)$ creates a particle of mass $M_1$ and a particle of mass $M_2-M_1$.
The propagator for the field $\phi_2(x)$ is 
\beq
D_2(E,p) = -i \, C_2^\prime
\left[\left(  \frac{p^2}{2M_2} - E \right)^{ 1/2}  - \frac{1}{\sqrt{2\mu a^2}} \right]^{-1}.
\label{prop2-a}
\eeq
This propagator has a pole at the energy
\beq
E_\mathrm{pole} = - \frac{1}{2 \mu a^2} + \frac{p^2}{2M_2}.
\label{Epole}
\eeq
The pole is associated with the bound state, which we will refer to as $X$.
The discontinuity of the propagator is
\bea
D_2(E+ i \epsilon,p) - D_2(E- i \epsilon,p) &=& 2 C_2^\prime\, 
\Bigg[  \frac{ \sqrt{E -  p^2/(2M_2)} }{ E -  p^2/(2M_2) + |\varepsilon_X|} \,  \theta\bigg(\! E -  \frac{p^2}{2M_n} \bigg)
\nonumber\\
&&\hspace{1.5cm} 
+  2\pi \sqrt{|\varepsilon_X|} \, \delta\bigg(\! E + |\varepsilon_X| -  \frac{p^2}{2M_2} \bigg) \Bigg].
\label{Discprop2a}
\eea
%

\begin{figure}[t]
\includegraphics[width=0.4\textwidth]{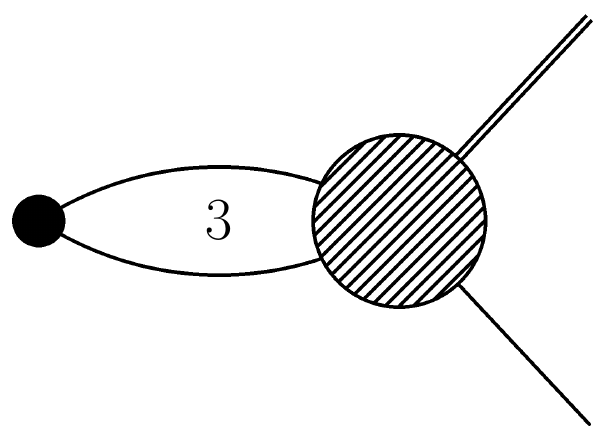}
\caption{
Amplitude for the creation of unparticle 3 at a point 
and its evolution to a loosely bound state of two particles and  a single particle with large relative momentum.
The bound state is represented by a double line.
}
\label{fig:pt3toX+1}
\end{figure}

The differential rate $dR$ for the production of either three particles or a single particle  plus $X$ can be obtained 
from the differential rate in Eq.~\eqref{dR-threshold1}
by taking into account the dependence of $ G_\mathrm{amp}(E_3-E_1,p)$ on $a$ and
replacing the discontinuity of the propagator $D_2(E_3-E_1,p)$ by the discontinuity given in Eq.~\eqref{Discprop2a}.
The amplitude for producing the bound state $X$ plus a single particle 
can be represented by the diagram in Fig.~\ref{fig:pt3toX+1}.
The integrated rate $R_X$ for producing  a single particle plus $X$ can be obtained
by using  the delta function in Eq.~\eqref{Discprop2a} to evaluate the integral over $\bm{p}$ in Eq.~\eqref{dR-threshold2}:
\beq
R_X = C_X  \sqrt{|\varepsilon_X|}
\big(E_3+ |\varepsilon_X|\big)^{1/2} 
\left| G_\mathrm{amp}\bigg(   \frac{M_1 E_3 - M_2|\varepsilon_X|}{M_3}, 
\big[2M_{12} \big(E_3+ |\varepsilon_X|\big) \big]^{1/2} \bigg)\right|^2,
\label{R-XD}
\eeq
where $C_X = C C_2^\prime (2M_{12})^{3/2}/\pi$.
The amputated 3-point function depends on $a$ explicitly through $\varepsilon_X$ in its two arguments
and also implicitly through the interactions between the particles.
In the limit $E_3 \gg |\varepsilon_X|$, its dependence on $E_3$ is given in Eq.~\eqref{<phi123>threshold}.
The dependence of $R_X$ on $E_3$ in that limit therefore has the power-law behavior 
\beq
R_X \longrightarrow C_X^\prime \sqrt{|\varepsilon_X|}\, 
E_3^{-\Delta_{12,3}+1/2},
\label{Rthreshold}
\eeq
where $C_X^\prime$ is a constant.
The exponent is $\Delta_3 - \Delta_2 - \Delta_1 + \tfrac1 2= \Delta_3-3$.

This case can be realized with neutral $D$ and $D^*$ mesons and the $X(3872)$~\cite{Braaten:2021iot}.
The $X(3872)$ is a resonance extremely close to the threshold in the $J^{PC}=1^{++}$ channel
of $D^{*0} \bar{D}^0$ and $D^0 \bar{D}^{*0}$.
The most precise measurements of the mass  by the  LHCb collaboration 
give an energy relative to the $D^{*0} \bar D^0$  threshold  of
$\varepsilon_X = -0.07 \pm 0.12$~MeV \cite{Aaij:2020qga,Aaij:2020xjx},
which implies $|\varepsilon_X| < 0.22$~MeV  at the 90\% confidence level.
This tiny binding energy corresponds to a large positive scattering length
$a=1/\sqrt{2\mu |\varepsilon_X|}$, where $\mu$ is the $D^{*0}\bar D^0$
reduced mass.
\begin{figure}[ht]
\centerline{
\includegraphics[width=0.75 \textwidth]{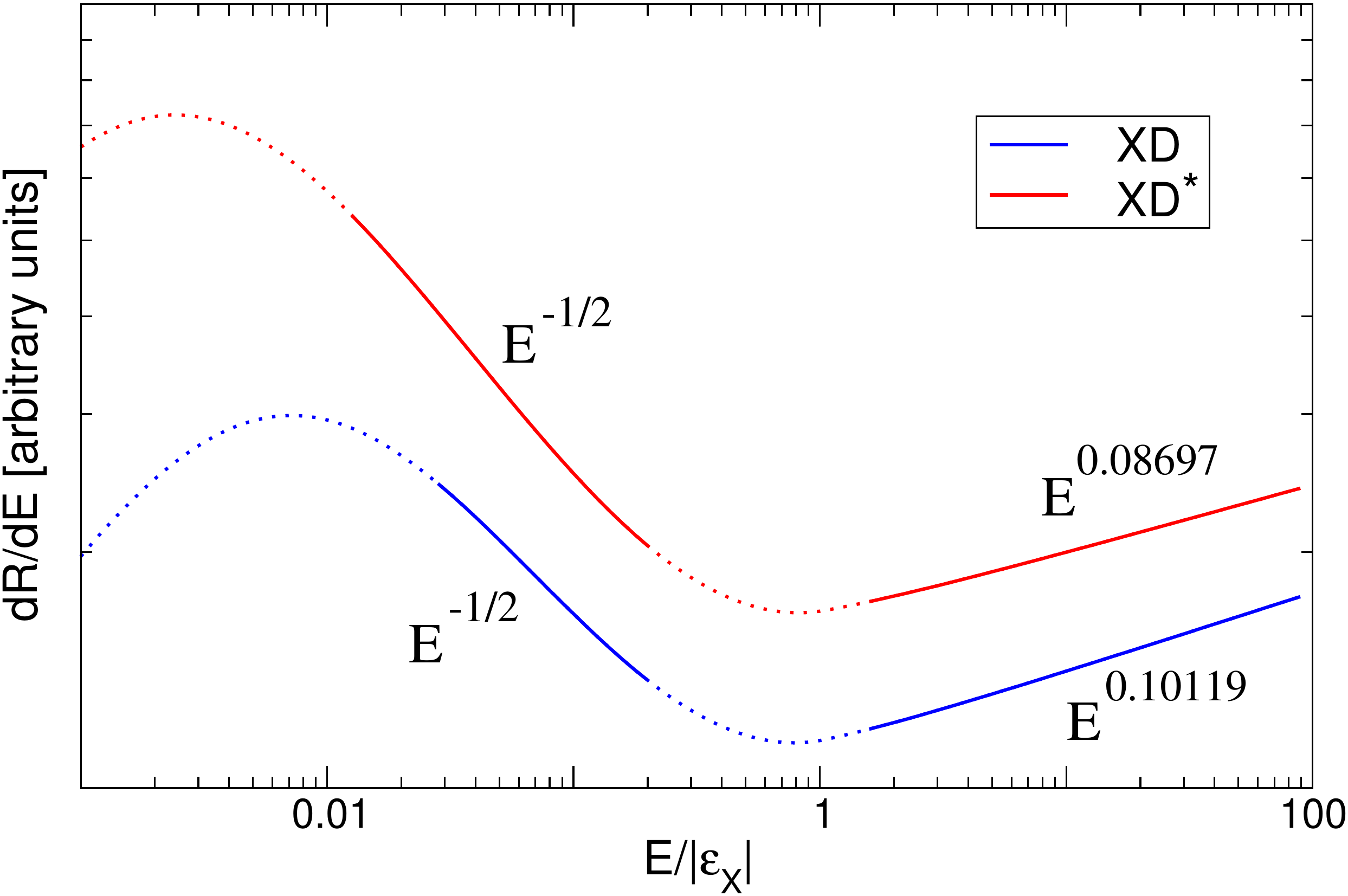}}
 \caption{
Production rates $dR/dE$ for $D^0X$ (solid curve) and $D^{*0}X$ (dashed curve)
 from the creation of neutral charm mesons at short distances as
 functions of the invariant energy $E$. The dotted lines show the
 transition between different scaling regions.
}
  \label{fig:XDvsE}
\end{figure}

The production rate $dR/dE$ of an $X$ and a neutral $D$ or $D^*$
meson near the threshold is determined by the energy scale
$\varepsilon_{DX} = 1/(2 \mu_{DX} a_{DX}^2)$,
where $\mu_{DX}$ is the reduced mass of the $XD$ (or $XD^*$) system
and $a_{DX}$ is the corresponding $S$-wave scattering length.
To leading order in the contact effective field theory
for the neutral $D$ and $D^*$ mesons, the
scattering lengths for $D^0X$ and $D^{*0}X$ are universal und
equal to $a$ multiplied
by a large negative coefficient: $a_{D^0X} = -9.7\, a$, $a_{D^{*0}X} = -16.6\, a$
\cite{Canham:2009zq}. Thus they show an additional enhancement beyond the
already large $D^{*0} \bar{D}^0$ scattering length $a$. Inserting explicit values
this leads to the tiny energy scales
$\varepsilon_{D^0X} =0.82$~keV or $\varepsilon_{D^{*0}X} =0.26$~keV.

As shown in Fig.~\ref{fig:XDvsE}, $dR/dE$ increases from zero at the
threshold to a peak near $\varepsilon_{DX}/|\varepsilon_X|\approx
0.01$ for $D^0 X$ and $0.004$ for $D^{*0} X$, respectively. 
Then it decreases to a local minimum at an energy of order
$|\varepsilon_X|$. In this region, the interior structure of the $X$
is not resolved and the scaling exponent $-1/2$ for two structureless
particles in the final state appears~\cite{Hammer:2021zxb}. Beyond the minimum,
there is a scaling region where $dR/dE$ increases with a power-law
determined by the three-body scaling dimensions $\Delta_3= 3.10119$ for
$D^0 X$ and $\Delta_3= 3.08697$ for $D^{*0} X$.
The power-law behavior of the amplitudes $\Gamma$ for  $D^0X$ and $D^{*0}X$
is predicted as $E^{0.10119}$ and $E^{0.08697}$,
respectively~\cite{Braaten:2021iot}. It has been
verified by explicit calculations of the point production rate
in a contact effective field theory
for the neutral $D$ and $D^*$ mesons.
There is a crossover to a more rapidly increasing production rate at
even higher energies when non-universal effects kick in, which
is not shown in Fig.~\ref{fig:XDvsE}.

\section{Conclusion}
\label{sec:conclusion}

 Nonrelativistic unparticles arise naturally in any system that can be described by a nonrelativistic field theory
close to a conformally invariant limit.
Systems of neutrons with small invariant energy created by point reactions are examples of 
nonrelativistic unparticles in nuclear physics \cite{Hammer:2021zxb}.
Systems of neutral charm mesons with small invariant energy created by point reactions 
are examples of nonrelativistic unparticles in particle physics~\cite{Braaten:2021iot}.
The concept of nonrelativistic unparticles is useful, because it allows scaling regions to be identified 
in which reaction rates have power-law behavior characterized by nontrivial exponents.
It would be interesting to find other physical realizations of nonrelativistic unparticles in nature.
It would also be interesting to exploit the remarkable control of interactions that is possible with ultracold atoms
to create new systems with unparticles.

We have used the analytic results for the three-point function for primary operators 
in a nonrelativistic conformal field theory 
in Refs.~\cite{Henkel:2003pu,Fuertes:2009ex,Volovich:2009yh} to derive the contribution
to the point production rate of an unparticle from its decay into another unparticle recoiling against a particle.
We also used it to derive a scaling law for the decay 
into a loosely bound 2-particle state recoiling against a particle with large momentum.
If analytic  results for the 4-point function of primary operators in a nonrelativistic conformal field theory were available, they could be applied to the elastic scattering of unparticles or the elastic scattering of an unparticle with a particle.
They could also be used to derive scaling laws for the elastic scattering
of a loosely bound 2-particle state and a particle at large momentum transfer.
It would be interesting to compare the analytic exponents with numerical results for 
the elastic scattering of $D^0$ and $D^{*0}$ with $X(3872)$ \cite{Braaten:2021iot}.

\acknowledgments
H.-W.H. acknowleges the hospitality of KITP, Santa Barbara
where part of this work was
carried out.
The research of E.B.\ was supported in part by the U.S.\ Department of Energy under grant DE-SC0011726.
H.-W.H. was supported in part by the National Science Foundation under
Grant No. NSF PHY-1748958,
by the Deutsche Forschungsgemeinschaft (DFG,
German Research Foundation) - Project-ID 279384907 - SFB 1245 and by the
German Federal Ministry of Education and Research (BMBF)
(Grant no.\ 05P21RDFNB).

\newpage



\newpage


 \end{document}